\documentclass[useAMS,usenatbib]{mn2e}
\usepackage{graphics,amsmath}
\begin{document}

\title[New Faint T Dwarfs in UKIDSS DR1]
{Four Faint T Dwarfs from the {\it UKIRT} Infrared Deep Sky Survey (UKIDSS) Southern Stripe }
\author[Chiu et al.\ ]{Kuenley Chiu$^1$, Michael C. Liu$^{2,3}$, Linhua Jiang$^4$, Katelyn N. Allers$^2$, Daniel P. Stark$^5$, 
\newauthor  Andrew Bunker$^{1,6}$, Xiaohui Fan$^4$, Karl Glazebrook$^7$, Trent J. Dupuy$^2$  \\
$^1$\,School of Physics, University of Exeter, Stocker Road, Exeter EX4 4QL, UK {\tt email: chiu@astro.ex.ac.uk}\\
$^2$\,Institute for Astronomy, University of Hawaii, 2680 Woodlawn Drive, Honolulu, HI 96822\\
$^3$\,Alfred P. Sloan Fellow\\
$^4$\,Steward Observatory, University of Arizona, 933 North Cherry Avenue, Tucson, AZ 85721\\
$^5$\,Department of Astronomy, California Institute of Technology, Mail Code 220-6, 1200 E. Calironia Blvd., Pasadena, CA 91125 \\
$^6$\,Anglo-Australian Observatory, Epping, NSW 1710, Australia\\
$^7$\,Centre for Astrophysics and Supercomputing, Swinburne University of Technology, Hawthorn, Victoria 3122, Australia\\
}
\date{accepted by MNRAS Letters, December 5, 2007}
\maketitle
\begin{abstract}
We present the optical and near-infrared photometry and spectroscopy of four faint T dwarfs newly discovered from the UKIDSS first data release.  The sample, drawn from an imaged area of $\sim$136 square degrees to a depth of $Y=19.9$ ($5\sigma$, Vega), is located in the SDSS Southern Equatorial Stripe, a region of significant future deep imaging potential.  We detail the selection and followup of these objects, three of which are spectroscopically confirmed brown dwarfs ranging from type T2.5 to T7.5, and one is photometrically identified as early T.   Their magnitudes range from $Y=19.01$ to 19.88 with derived distances from 34 to 98 pc, making these among the coldest and faintest brown dwarfs known.  The sample brings the total number of T dwarfs found or confirmed by UKIDSS data in this region to nine, and we discuss the projected numbers of dwarfs in the future survey data.  We estimate that $\sim240$ early- and late-T dwarfs are discoverable in the UKIDSS LAS data, falling significantly short of published model projections and suggesting that IMFs and/or birthrates may be at the low end of possible models.   Thus, deeper optical data has good potential to exploit the UKIDSS survey depth more fully, but may still find the potential Y dwarf sample to be extremely rare. 
\end{abstract}
\begin{keywords}

surveys -- catalogues -- Stars: brown dwarfs -- techniques: photometric -- techniques: spectroscopic -- Infrared: Stars
\end{keywords}

\section{Introduction}
\label{sec:INTRO}

\begin{figure*}
\resizebox{0.65\textwidth}{!}{\includegraphics{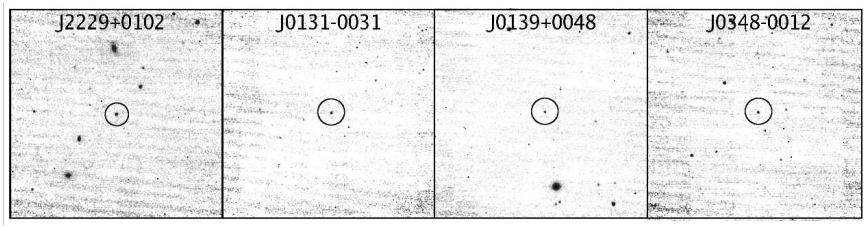}}
\caption{Finding charts for the four T dwarfs discovered in this work.  Frames are from {\it UKIRT} UFTI followup imaging in $J$-band.  Images have been Gaussian smoothed by 2 pixels to enhance features.  Each panel is $2'\times2'$, oriented N up, E left.  
\label{fcsfig}}
\end{figure*}

Members of the coldest brown dwarf spectral type currently observed, the ``T'' class \citep{kirkpatrick05}, are now being discovered quite regularly thanks to a growing archive of optical and near-infrared survey data.  As a result, this sample of once-rare objects has increased dramatically from the handful known only a few years ago \citep{nakajima95, burgasser99, cuby99, strauss99}, to the dozens gradually retrieved by searches through dedicated surveys \citep{burgasser02,geballe02, knapp04, chiu06, stern07}, and ultimately towards what is predicted to be the bulk discovery of hundreds in the next few years \citep{deacon06}.  As the growth of this discovered population continues, the study of T dwarfs has also naturally expanded to ever more detailed examination of their physical properties, and has fueled observations of rarer, but highly interesting properties, such as binarity, {\it e.g.} \citet{liu06}.  
 
The advance in the discovery rate of T dwarfs has been a direct product of the coupling of optical and near-infrared survey techniques.  With the development of the {\it UKIRT} Infrared Deep Sky Survey \citep[UKIDSS,][]{lawrence07}, the opportunity for automated candidate identification has grown enormously.  The first T dwarf discoveries from UKIDSS are confirming the survey's powerful red object finding capability  \citep[e.g.,][]{kendall07,lodieu07,warren07b}.

In this paper, we present the first results of our ongoing search for the reddest objects in the  UKIDSS data: the coolest brown dwarfs and very high redshift quasars.  We have found three spectroscopically confirmed early- to late-T dwarfs in the UKIDSS DR1 area of approximately 150 square degrees, and one strong photometrically/spectroscopically identified early-T candidate, and we describe the techniques used to identify these objects, their followup, and characterization.  We also compare the sample with other discovered brown dwarfs in this area, and discuss predictions of brown dwarf discovery rates for the full UKIDSS project.  In this paper, all near-IR magnitudes are expressed in the UKIDSS Vega-normalized system. (note that the $Y$ band photometry used here does not include a possibly necessary correction of 0.1 mag; see Warren et al. 2007a for details). However, accompanying photometry in the optical $i/z$ bands are on the AB system, as used by the SDSS project \citep{fukugita96,lupton99}. 

\section{Candidate Selection}
\label{sec:datasources}
This work is based on the UKIDSS first data release (DR1), and was aimed at mining the Southern Equatorial Stripe, pioneered by the SDSS (Stripe 82). The LAS (Large Area Survey), the wide and shallow sub-survey of UKIDSS, imaged this area partially over the 2005B semester, covering approximately 136 square degrees to a depth of [$Y,J,H,K$]=[20.2, 19.6, 18.8, 18.2] ($5\sigma$, Vega).   
\begin{table*}  
 \centering 
 \begin{minipage}{190mm}
  \caption{Observed T dwarf properties.  Optical $iz$ photometry was obtained with the {\it SOAR}/SOI instrument, and $YJHK$ photometry was obtained from the UKIDSS database, or {\it UKIRT}/UFTI and {\it AAT}/IRIS2 observations as indicated. \label{phottable}}
  \begin{tabular}{@{}lcccccccl   @{}}
  \hline
Object ID     & $i$  &  $z$  &  $Y$    &  $J$     &  $H$     &  $K$	& Type & Source \\
 ULAS (J2000)      & (AB) &  (AB) & (Vega)& (Vega) & (Vega) & (Vega) &		   \\

\hline
J$222958.30+010217.2$  & $24.38\pm0.48$ & $21.16\pm0.06$ & $19.01\pm0.06$  &  $17.87\pm0.03$  &  $17.48\pm0.05$  &  $17.21\pm0.09$ & T2.5    & UKIDSS	  \\
   		     &  	      & 	       &		 &  $17.88\pm0.02$  &  $17.58\pm0.02$  &		 &	   & UFTI  \\

J$013117.53-003128.6$ & $>24.5$        & $22.78\pm0.19$ & $19.88\pm0.21$  &  $18.78\pm0.12$  &  $18.61\pm0.15$  &  ---		 & T3.0  & UKIDSS   \\
   		     &  	      & 		&		 &  $18.60\pm0.04$  &  $18.48\pm0.04$  &  $18.41\pm0.04$ &	   & UFTI, IRIS2 \\

J$013939.77+004813.8$  & $>25.0$        & $22.55\pm0.20$ & $19.57\pm0.13$  &  $18.69\pm0.09$  &  $18.61\pm0.17$  &  ---		 & T7.5    & UKIDSS   \\
		     &  	      & 	       &		 &  $18.43\pm0.04$  &  $19.12\pm0.05$  &  $19.23\pm0.06$ &	   & UFTI, IRIS2\\

J$034832.97-001258.3$ & $>24.8$        & $21.56\pm0.12$ & $19.73\pm0.11$  &  $18.89\pm0.08$  &  ---	       &  ---		 & earlyT & UKDISS    \\
		     &  	      & 	       &		 &  $18.72\pm0.04$  &  $17.95\pm0.02$  &		 &	   & UFTI \\
	 
\hline		
\end{tabular}							    
\end{minipage}	
\end{table*}								      	 	     	 								 \begin{figure*}
\resizebox{0.6\textwidth}{!}{\includegraphics{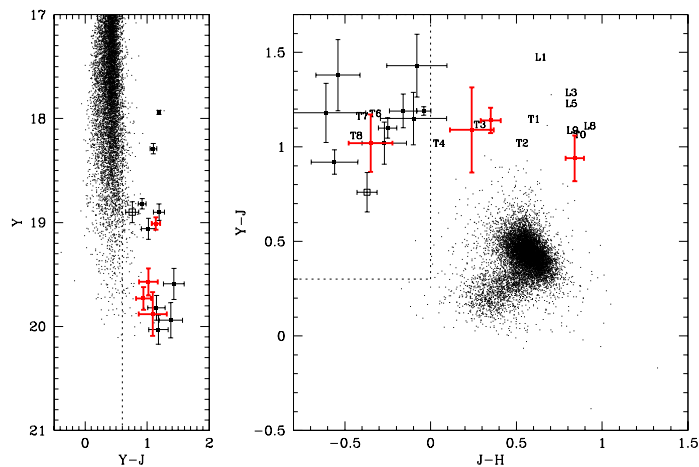}}
\vspace*{-1.6in}
\caption{The four faint T dwarfs discovered in this work are plotted (red points with thick error bars) in $(Y-J)$ versus $Y$ color-magnitude (left) and $YJH$ color-color (right) diagrams.   Normal stars are shown as small points, and the concurrently discovered objects of \citet{lodieu07} are shown for comparison (black points with light error bars). The extremely late-type T8.5 dwarf of \citet{warren07b} is the open square with error bars.    The sample of Lodieu et al. includes: four new T dwarfs and one previously known SDSS T dwarf found in the same area as our sample, and 4 T dwarfs discovered in additional area of the UKIDSS DR1 not covered in the present work.     In the left panel, the dotted line indicates the typical $Y-J>0.6$ selection cut used to identify dwarf candidates.   In the right panel, the dotted rectangular region indicates the range of predicted colors for late T and Y dwarfs   \citep{leggett04}.   Small text labels follow the approximate locus of brown dwarf example types, selected from Table 10 of \citet{hewett06}.   The normal stellar sample was retrieved from a match of unidentified point sources appearing in both UKIDSS and SDSS.  
\label{colorfig}}
\end{figure*}		

For the science user seeking the reddest objects, the most important attribute of the UKIDSS database is the  matching of SDSS photometry with LAS-detected objects (Hambly et al., Irwin et al., in prep).  This non-trivial cross-matching feature, operated on $N\sim10^8$ objects, allows identification of sources displaying flux decrements of interest, and most importantly for this work, across the optical and near-IR bands: {\it e.g.} brown dwarfs and high-redshift quasars. 
 
Using the merged UKIDSS/SDSS photometry, we ranked LAS-detected point-source objects with a priority based on several factors.   The UKIDSS $Y$ and/or $J$ magnitudes were sorted, giving emphasis in followup observations to the brightest candidates first, typically down to a magnitude limit of $Y=19.9$ dictated by the feasibility of spectroscopic observations.  Photometric errors were examined to measure the confidence of weak detections given the varying depth of exposures.  The $i-Y$ color then yielded the primary ranking, with high values indicating candidates with the greatest potential.  A limit of $i-Y>3.5$ was imposed to produce managable candidate numbers, and when a UKIDSS point source lacked an SDSS source within several arcseconds, a lower limit on the $i-Y$ value was calculated, generally implying a large and interesting but uncertain $i-Y$ value.  The nominal SDSS flux limits reach $[u,g,r,i,z]=[22.0,22.2,22.2,21.3,20.5]$ ($5\sigma$, AB).  

Imposing the $i-Y$ and magnitude criteria, the surface density of candidates from the database was $\sim 5$ per square degree.  Candidates were then subjected to automated as well as visual inspection against their archived imaging to check for defects, among which cosmic rays, interpolation artifacts, cross-talk and persistence features are the most common in the UKIDSS and SDSS imagery.  In addition, deeper imaging data from repeated scans of the SDSS were used to verify photometry and color decrement in $i-z$.  In some areas, where 10 or more SDSS runs contributed to the deeper imaging, this allowed verification to 1.25 mag deeper than the normal SDSS survey depth, as has been demonstrated in a few searches to date \citep{jiang06,jiang07}. 
\section{Observations}
The preliminary inspection procedures reduced candidate numbers dramatically, and a more restricted list of good objects, numbering $\sim 1$ per square degree, was subjected to followup imaging.  This was undertaken to measure the faint $i/z$ flux of the objects, and to confirm the near-IR photometry of the new WFCAM images.  As detailed in Table \ref{phottable}, these observations were taken at several facilities: the UFTI instrument (at {\it UKIRT}), IRIS2 ({\it AAT}), and SOI ({\it SOAR}).  Finding charts for the four T dwarfs, based on the UFTI $J$ imaging, are displayed in Figure \ref{fcsfig}.   Typical exposure times in the near-IR and optical were 300 seconds.   Based on these imaging trials and other work with larger samples \citep[{\it e.g.},][]{chiu07}, the UKIDSS photometry is found to be relatively reliable at the $>5\sigma$ detection level, as long as images are not affected by artifacts.  However, some larger discrepancies were encountered, such as with the $H$-band measurements of J0139+0048.
 
Following imaging verification, 20 objects continued to display interesting colors, including two T dwarfs later spectroscopically observed in collaboration with \citet{lodieu07} and \citet{warren07b} -- ULAS J2239+0032 (T5.5) and J0034--0052 (T8.5).   In addition, among these objects was the $z=5.86$ quasar confirmed by \citet{venemans07}.  Optical spectroscopy was then carried out on the remaining candidates using the EMMI instrument at the ESO 3.6m {\it NTT} telescope in the $R\sim600$ RILD mode (program 078.A-0398), or the MARS red spectrograph at KPNO ($R\sim450$), both covering $\sim5500-9500$\AA.   Candidates typically received $3\times15$ minutes exposure each, and the MARS nod-and-shuffle mode was employed to improve sky line background subtraction in the troublesome $\lambda>8000$ \AA ~region.  

The optical spectroscopy resulted in the preliminary identification of five L dwarfs and four T dwarfs.  The T dwarfs are plotted in color and magnitude in Figure \ref{colorfig}, showing that the sample extends to the range of the faintest late T dwarfs yet known.  In addition, we note that the $i-Y$ selection technique employed here is able to discover dwarfs outside the near-infrared selection box shown, which otherwise relies on the presence of flux in all of the $YJH$ bands.  The relatively well-separated dwarf color locus versus type allowed rough classification of discovered candidates based on photometry, and this is overlaid in Figure \ref{colorfig}.  

\begin{figure}
\resizebox{0.45\textwidth}{!}{\includegraphics{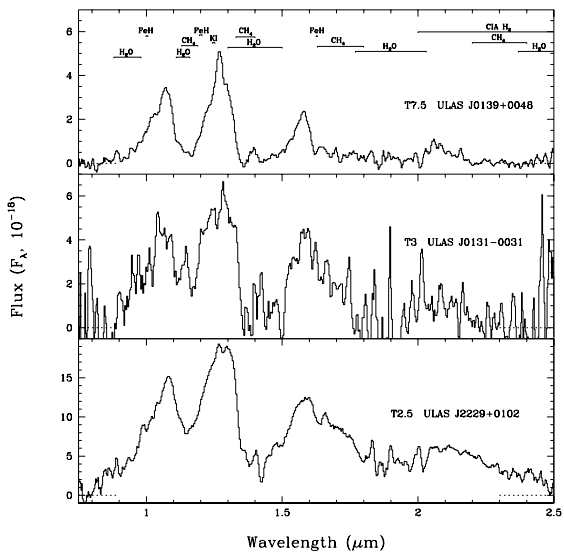}}
\vspace*{-0.2in}
\caption{Near-infrared spectra of the 3 T dwarfs confirmed in this work.  Molecular absorption bands are indicated in the top panel, and observation details are discussed in Section 3. }
\label{spectrafig}
\end{figure}

To confirm the T dwarf types, we obtained low-resolution ($R\approx$100) near-IR spectra on UT20061118 using the NASA {\it IRTF}.   Conditions were photometric with excellent seeing at $\sim0\farcs5$ FWHM at $K$-band near zenith.  We used the SpeX spectrograph \citep{rayner98} in prism mode, obtaining 0.8--2.5~\micron\ spectra in a single order, with the $0\farcs8$ slit oriented at the parallactic angle.  Objects were nodded along the slit in an ABBA pattern with individual exposure times of 180~sec, and observed over airmass  1.1--1.3.  The typical on-source exposure time was 54 min, and A0~V stars were observed for flux and telluric calibration.  All spectra were reduced using the SpeXtool software package \citep{vacca03,cushing04}, and are shown in Figure \ref{spectrafig}.  To our knowledge, these are the faintest spectra yet published from the {\it IRTF}/SpeX. 

\begin{figure}
\resizebox{0.5\textwidth}{!}{\includegraphics{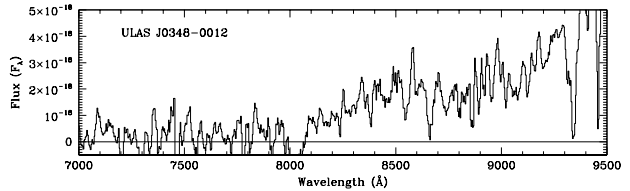}}
\vspace*{-2.5in}
\caption{Optical spectrum of ULAS J0348-0012, determined to be a T dwarf from optical/near-IR colors and the lack of quasar emission features from $7000-9500$\AA.  The spectrum has been smoothed by 5 pixels.
}
\label{0348fig}
\end{figure}

The spectral types of the T dwarfs were determined by matching to template dwarf spectra \citep{burgasser06a}, and have an uncertainty of 0.5 in type.  Object J0348-0012 lacks a near-IR spectrum, but is most likely an early T dwarf based on the absence of quasar emission features in the optical spectrum (Figure \ref{0348fig}), and its photometry as shown in Figure 2.  Using the spectral types of the three spectroscopically confirmed Ts, the photometric distances to each dwarf were derived from the relation of  \citet{liu06}, and are presented in Table \ref{disttable}.

We also undertook high angular resolution imaging of dwarf J0139+0048.  Only six dwarfs of spectral type T7.5 or cooler have been imaged before at high angular resolution, with either {\it HST} \citep[e.g.][]{burgasser06b} or Keck LGS (Liu et al., in prep).  We imaged J0139+0048 on UT20070906 using the sodium LGS adaptive optics system of Keck II.  Conditions were photometric with excellent seeing and we used the NIRC2 camera to obtain dithered $K$-band images, with a total on-source integration time of 720~seconds.  The resulting data have FWHM of 0.050$\pm$0.03\arcsec\ and Strehl ratios of 41$\pm$11\%.  The object appears single in our images, with no companions at 0.1--2.0\arcsec\ separation up to 3.5~mags fainter ($K=22.7$).  At 35~pc and an assumed age of 1--5~Gyr, this corresponds to companions of 12--30~$M_J$, based on the models of \citet{baraffe03}.

\begin{table}  
 \centering 
 \begin{minipage}{3.4in}
  \caption{Photometric distances of the spectroscopically confirmed T dwarf sample. calculated based on each of the average $JHK$ magnitudes and the spectral type to absolute magnitude relation of \citet{liu06}.  The first series of $JHK$ distances correspond to the ``excluding known binaries fit'', and the second series correspond to ``excluding known and possible binaries''.  Distances are in parsec. \label{disttable}}
  \begin{tabular}{@{}lccccccc   @{}}
  \hline
Object     & Type  &   $d_{J1}$ &  $d_{H1}$ &  $d_{K1}$ &  $d_{J2}$ &  $d_{H2}$ &  $d_{K2}$ \\
 \hline
J2229+0102 &  T2.5 & 63.3  &  65.7  & 60.8  & 47.2 & 46.9  &  42.6 \\
J0131--0031 & T3.0 & 89.3  &  97.8  & 97.4  & 67.2 & 70.8  &  69.5 \\
J0139+0048 &  T7.5 & 34.1  &  32.9  & 36.9  & 35.1 & 33.5  &  38.1 \\

\hline		
\end{tabular}							    
\end{minipage}	
\end{table}								      	 	     					     	 		

\section{T Dwarf Sky Density and UKIDSS Discovery Rate}
These four new faint T dwarfs were drawn from a searched area defined by the overlap of the UKIDSS DR1 dataset with the SDSS Southern Equatorial Stripe, which ranges from RA $\sim$22$^h$ to 4$^h$, and Dec $=\pm1.265^\circ$.   We calculate this overlap to be approximately 136 square degrees by rectangular approximation using a 0.2 degree mesh, but with uncertainty of $\pm$10\% due to incomplete UKIDSS tiling and areas lost due to extremely bright stars and other normal imaging defects.  In this area, collaborators concurrently identified four other new T dwarfs ranging from type T4.5 to T8.5, which we also selected during this search, but did not pursue independently \citep[see][]{lodieu07,warren07b}.  One other T dwarf (T4, SDSS J0207+0000) was reconfirmed, having been found previously in the SDSS data alone \citep{geballe02}.  Including our present sample then, this brings the total number of T dwarfs to nine in this area, down to a searched magnitude limit of $Y=19.9$.  

A recent estimate of T dwarf numbers in the UKIDSS LAS complete area is provided by \citet{deacon06} (hereafter DH06), based on a projection of 4000 square degrees reaching a depth of $[Y,J,H,K]=[20.5,20.0,18.8,18.4]$.  Aside from the increased area, this differs from the present DR1 flux limits by an increase of 0.4 mag in depth of the $J$ imaging ($J=19.6$, $5\sigma$, Warren et al. 2007a).  This is because the LAS $J$ band is planned to have a second epoch of imaging to aid determination of proper motions, among other goals.  Also, their assumed limiting $Y$ magnitude is slightly deeper than that actually achieved by the survey ($Y=20.2$, $5\sigma$).  With these flux limits, and depending on various assumed initial mass functions (IMF) and birthrates, the DH06 discovery estimates for the UKIDSS LAS survey are from 321 to 1375 early T dwarfs, 657 to 2014 late T dwarfs, and 25 to 100 Y dwarfs, which Deacon and Hambly operationally define as objects with $T<770K$.  At present, the actual spectral criteria and $T_{eff}$ that would warrant a new type are unknown, though the presence of NH$_3$ absorption features is expected to be important \citep{leggett07}.  

Based on these nine T dwarfs found to date in the Southern Equatorial Stripe, we suggest that the eventual numbers of discovered dwarfs will not be able to reach the theoretical sky density of DH06, even after correcting for different assumptions about the increased expected depth of the  full UKIDSS  imaging.  The nine T dwarfs found in 136 square degrees suggest that by simple area scaling, approximately 264 dwarfs in the {\it totaled} early- to late-T classes may be found in 4000 square degrees of imaging to these depths (searched depth of $Y=19.9$), approximately an order of magnitude less than the typical numbers expected by DH06.  

The difference in quoted survey limits must also be taken into account to correctly compare the samples.  However, after doing so, a significant shortfall remains.  Increasing the assumed searched depth from $Y=19.9$ (actually searched in this sample) to $Y=20.5$ (the flux limit of DH06), under an assumption of uniformly distributed objects, this would improve the limiting flux to $1/1.73\times$ the present value, the probed distance to $1.31\times$, and therefore the searched spherical volume to $2.29\times$ the present value -- 606 early+late T dwarfs.  However, this still represents an approximately $3.5\times$ deficiency compared to the average DH06 projections.  In addition, if the expected vertical scale height for dwarfs were to differ from normal stars (such as a more rapid falloff), this could decrease the discovered numbers.  
 
We note that while a future increase in survey $J$ depth would increase the certainty of selected candidates through higher $S/N$ detections, it would not increase the absolute numbers of discovered dwarfs as dramatically as the factor of $8\times$ suggested by \citet{lodieu07}, because the colors of T dwarfs are expected (and selected) to display $Y-J>0.5$.  As a result, in the current sample being searched to $Y=19.9$, the survey depth of $J=19.6$ has not been a limiting factor.  Similarly, for the DH06 survey depth assumption of $Y=20.5$ (which is corrected for above), the simultaneous $J=20.0$ depth would not be limiting, and therefore the difference in survey $Y$ depth can only account for one-sixth to one-half of the large discrepancy between the discovered dwarfs versus the DH06 predicted sky density.  This shortfall, if affirmed as the UKIDSS dwarf census grows, may imply that the actual IMF/birthrate of early and late T dwarfs falls at the very low end of DH06 model scenarios, and that Y dwarfs (as defined by DH06) will be proportionally even more rare.  The T8.5 dwarf (J0034--0052) of Warren et al. has an estimated $T_{eff}$ of 650--750$K$ but with spectral characteristics only slightly different from previously identified late-T dwarfs.  Y dwarfs with NH$_3$ absorption features may thus be rarer even than the 25--100 predicted by DH06 in the full UKIDSS LAS -- as challenging to find as $z>6$ quasars. 

\section{Conclusion}
We have conducted a search for high-redshift quasars and the coldest brown dwarfs in $\sim136$ square degrees of the UKIDSS DR1 in the Southern Equatorial Stripe (SDSS Stripe 82).  Searching the combined UKIDSS and SDSS databases, followed by dedicated imaging and spectroscopy, has yielded four newly discovered T dwarfs, three of which are spectroscopically identified and one is strongly inferred based on near-IR photometry and optical spectroscopy.  These are in addition to the four T dwarfs discovered independently by collaborators and one SDSS T dwarf previously known in this area.  The four discovered dwarfs are among the faintest and coldest dwarfs found to date.  From this ongoing search, we expect approximately $\sim264$ early+late T dwarfs may be found in the total 4000 square degrees of the final UKIDSS data.  As deep optical data from SDSS become available, this search will continue to seek fainter and cooler dwarf types, in hopes of the still elusive Y dwarf class, and an ever increasing number of these once rare objects. 

\section{Acknowledgments} KC acknowledges support from a PPARC rolling grant, MCL acknowledges NSF grants AST-0407441, AST-0507833 and an Alfred P. Sloan Research Fellowship, and AJB acknowledges support of a Philip Leverhulme Prize.  MCL and KNA are Visiting Astronomers at the IRTF, which is operated by the Univ. of Hawaii under NCC 5-538 for NASA.  We also thank Richard Ellis, Daniel Stern, and Johan Richard for generous assistance with observations contributing to the identification of the sample here.  We are grateful to the UKIDSS collaboration for making available the high quality imaging and catalog products to the community.  We appreciate the expert assistance of the staffs of the IRTF, NTT, AAT, SOAR, KPNO, and UKIRT observatories during this research.


\begin{thebibliography}{}
\bibitem[Abazajian et al.(2004)]{abz} Abazajian, K., et al. 2004, AJ, 128, 502
\bibitem[Baraffe et al.(2003)]{baraffe03} Baraffe, I., et al. 2003, A\&A, 402, 701
\bibitem[Burgasser et al.(1999)]{burgasser99} Burgasser, A.~J., et al.\ 1999, ApJl, 522, L65 
\bibitem[Burgasser et al.(2002)]{burgasser02} Burgasser, A.~J., et al.\ 2002, ApJ, 564, 421 
\bibitem[Burgasser et al.(2006a)]{burgasser06a} Burgasser, A.~J., Geballe, T.~R., Leggett, S.~K., Kirkpatrick, J.~D., \& Golimowski, D.~A.\ 2006a, ApJ, 637, 1067 
\bibitem[Burgasser et al.(2006b)]{burgasser06b} Burgasser, A.~J., et al.\ 2006b, ApJs, 166, 585
\bibitem[Chiu et al.(2005)]{chiu05} Chiu, K., et al.\ 2005, AJ, 130, 13 
\bibitem[Chiu et al.(2006)]{chiu06} Chiu, K., et al.\ 2006, AJ, 131, 2722 
\bibitem[Chiu et al.(2007)]{chiu07} Chiu, K., Richards, G.~T., Hewett, P.~C., \& Maddox, N.\ 2007, MNRAS, 375, 1180 
\bibitem[Cuby et al.(1999)]{cuby99} Cuby, J.~G., et al.\ 1999, A\&A, 349, L41 
\bibitem[Cushing et al.(2004)]{cushing04} Cushing, M.~C., Vacca, W.~D., \& Rayner, J.~T.\ 2004, PASP, 116, 362 
\bibitem[Deacon \& Hambly(2006)]{deacon06} Deacon, N.~R., \& Hambly, N.~C.\ 2006, MNRAS, 371, 1722 
\bibitem[Fukugita et al.(1996)]{fukugita96} Fukugita, M., et al.\ 1996, AJ, 111, 1748 
\bibitem[Geballe et al.(2002)]{geballe02} Geballe, T.~R., et al.\ 2002, ApJ, 564, 466 
\bibitem[Hewett et al.(2006)]{hewett06} Hewett, P.~C., Warren, S.~J., Leggett, S.~K., \& Hodgkin, S.~T.\ 2006, MNRAS, 367, 454 
\bibitem[Jiang et al.(2006)]{jiang06} Jiang, L., et al.\ 2006, AJ, 131, 2788 
\bibitem[Jiang et al.(2007)]{jiang07} Jiang, L., et al.\ 2007, ArXiv e-prints, 708, arXiv:0708.2578 
\bibitem[Kendall et al.(2007)]{kendall07} Kendall, T.~R., et al.\ 2007, A\&A, 466, 1059 
\bibitem[Kirkpatrick(2005)]{kirkpatrick05} Kirkpatrick, J.~D.\ 2005, ARAA, 43, 195 
\bibitem[Knapp et al.(2004)]{knapp04} Knapp, G.~R., et al.\ 2004, AJ, 127, 3553 
\bibitem[Lawrence et al.(2007)]{lawrence07} Lawrence, A., et al.\ 2007, MNRAS, 379, 1599 
\bibitem[Leggett et al.(2004)]{leggett04} Leggett, S.~K., Allard, F., Burgasser, A.~J., Jones, H.~R.~A., Marley, M.~S., \& Tsuji, T.\ 2004, astro-ph/0409389 
\bibitem[Leggett et al.(2007)]{leggett07} Leggett, S.~K., et al.\ 2007, ApJ, 667, 537 
\bibitem[Liu et al.(2006)]{liu06} Liu, M.~C., et al.\ 2006, ApJ, 647, 1393 
\bibitem[Lodieu et al.(2007)]{lodieu07} Lodieu, N., et al.\ 2007, MNRAS, 379, 1423 
\bibitem[Lupton et al.(1999)]{lupton99} Lupton, R.~H., Gunn, J.~E., \& Szalay, A.~S.\ 1999, AJ, 118, 1406 
\bibitem[Nakajima et al.(1995)]{nakajima95} Nakajima, T., et al.\ 1995, Nature, 378, 463 
\bibitem[Rayner et al.(1998)]{rayner98} Rayner, J.~T., et al.\ 1998, Proc. SPIE, 3354, 468 
\bibitem[Stern et al.(2007)]{stern07} Stern, D., et al.\ 2007, ApJ, 663, 677 
\bibitem[Strauss et al.(1999)]{strauss99} Strauss, M.~A., et al.\ 1999, ApJl, 522, L61 
\bibitem[Vacca et al.(2003)]{vacca03} Vacca, W.~D., Cushing, M.~C., \& Rayner, J.~T.\ 2003, PASP, 115, 389 
\bibitem[Venemans et al.(2007)]{venemans07} Venemans, B.~P., et al.\ 2007, MNRAS, 376, L76 
\bibitem[Warren \& Hewett(2002)]{warren02} Warren, S., \& Hewett, P.\ 2002, ASP Conf.~Ser.~283: A New Era in Cosmology, 283, 369 
\bibitem[Warren et al.(2007a)]{warren07a} Warren, S.~J., et al.\ 2007a, MNRAS, 375, 213 
\bibitem[Warren et al.(2007b)]{warren07b} Warren, S.~J., et al.\ 2007b, ArXiv e-prints, 708, arXiv:0708.0655 
\end{thebibliography}
\end{document}